\documentclass[11pt]{article}
\usepackage{moriond,epsfig}
\usepackage[latin1]{inputenc}
\usepackage[OT1]{fontenc}

\def\be{\begin{equation}}
\def\ee{\end{equation}}
\def\bea{\begin{eqnarray}}
\def\eea{\end{eqnarray}}

\begin{document}
\vspace*{4cm}
\title{MACH'S PRINCIPLE: EXACT FRAME-DRAGGING \\
BY ENERGY CURRENTS IN THE UNIVERSE}

\author{CHRISTOPH SCHMID}

\address{Institute of Theoretical Physics, ETH Zurich\\
CH-8093 Zurich, Switzerland}

\maketitle\abstracts{We show that the dragging 
of axis directions of local inertial frames 
by a weighted average of the energy currents in the universe 
(Mach's postulate) is exact for all linear perturbations 
of all Friedmann-Robertson-Walker universes and for all types of matter.}


\section{Mach's Principle}

\subsection{The Observational Fact: 'Mach zero'}
\label{subsec:Mach.zero}


The time-evolution of local inertial axes,
i.e. the local 
    {\it non-rotating frame}
is
    {\it experimentally} 
determined by the spin axes of 
    {\it gyroscopes,}
as in inertial guidance systems in airplanes and satellites.
This is true both in Newtonian physics (Foucault 1852) 
and in General Relativity.


It is an  
    {\it observational fact}
within present-day accuracy that the
    {\it spin axes of gyroscopes} 
do not precess 
    {\it relative to quasars}.
This observational fact has been named 'Mach zero',
where 'zero' designates that this fact is not yet Mach's principle,
it is just the observational starting point.~---
There is an extremely small dragging effect 
by the rotating Earth on the spin axes of gyroscopes,
the Lense - Thirring  effect,
which makes the spin axes of gyroscopes precess relative to quasars 
by 43 milli-arc-sec per year. 
It is hoped that one will be able to detect this effect 
by further analysis of the data which have been taken by Gravity Probe B.


\subsection{The Question}


    {\it What physical cause}
explains the observational fact 'Mach zero' ?
Equivalently: What physical cause determines 
the time-evolution of gyroscope axes ?     
In the words of John A. Wheeler: 
    {\it Who gives the marching orders} 
to the spin axes of gyroscopes, i.e. to inertial axes ?


\subsection{Mach's Postulate}


An answer to this fundamental question was formulated by Ernst Mach
in his postulate (1883) 
that inertial axes (i.e. the spin axes of gyroscopes)
    {\it exactly follow} 
an average of the motion of the 
    {\it masses in the universe:} 
Mach postulated 
    {\it exact frame dragging}
of inertial axes by the motion of cosmological masses,
    {\it not merely a little bit} 
of frame-dragging as in Lense-Thirring effect. 
 

Mach did not know,  
    {\it what mechanism}, 
what new force could do the job,
he merely stated: 
'the laws of motion could be conceived ...'.
Mach also asked: ``{\it What share} has every mass 
in the determination of direction ... in the law of inertia?
No definite answer can be given by our experiences.''


\subsection{Our Results}


We have shown that 
     {\it exact dragging} 
of inertial axis directions, i.e. Mach's Principle,
follows from Cosmological General Relativity for 
     {\it general, linear perturbations} 
of FRW backgrounds with $K = (\pm 1, 0).$ 
This also holds for FRW backgrounds with arbitrarily small 
energy density and pressure compared to $\rho_{\rm crit}$ 
(Milne limit of FRW universe).

These results have been demonstrated for the first time 
         in our paper~\cite{CS.I} 
for $K=0,$ 
     and in our paper~\cite{CS.II}
for $ K = (\pm 1, 0).$


\section{Theoretical Results and Tools}

\subsection{Cosmological Vorticity Perturbations}
\label{Vorticity.perturbations}


The vector sector of cosmological perturbations
is the sector of vorticity perturbations. 
Two important theorems 
for the vorticity sector are needed to understand the following summary:

\begin{enumerate}
\item
  The  
      {\it slicing} 
  of space-time in slices $\Sigma_t$ of fixed time is
      {\it unique}. 
  The 
      {\it lapse} 
  function (elapsed measured time between slices) 
  and $g_{00}$ are 
      {\it unperturbed}. 
\item 
  The 
      {\it intrinsic geometry} 
  of each slice $\Sigma_t,$ i.e. of 3-space, remains
      {\it unperturbed.} 
\end{enumerate}
The 
      {\it coordinate choice uniquely adapted} 
to our 3-geometry 
is comoving Cartesian coordinates for FRW with $K=0,$
resp. comoving spherical coordinates for $K = (\pm 1, 0).$
Hence the only quantity referring to vorticity perturbations is the
      {\it shift} 
3-vector  $\beta^i$ (resp $\beta_i = g_{0i}$):
\begin{eqnarray}
ds^2 &=& - dt^2 + a^2 
[\, d\chi^2 + R^2_{\rm com} ( d\theta^2 + \sin^2\theta \, d\phi^2  ) ] 
+ 2  \beta_i \, dx^i dt,
\\
R_{\rm com} &=& (\chi, \, \sin \chi, \, \sinh \chi).
\end{eqnarray}
%


\subsection{Gravitomagnetism}


The general     
     {\it operational definitions}
of the gravitomagnetic and gravitoelectric fields 
are given via measurements by 
     {\it FIDO}s ({\it Fid}ucial {\it O}bservers)
with 
     {\it LONB}s 
    ({\it L}ocal {\it O}rtho-{\it N}ormal {\it B}ases),
where LONB components are denoted by 
     {\it hats} 
over indices.

     {\it Gravitoelectric field} 
$\vec{E}_{\rm g} \equiv \vec{g} :$

\begin{equation} 
\frac{d}{dt} \, p_{\, \hat{i}} 
\equiv m \, E_{\, \hat{i}}^{\rm g} \quad
\mbox{free-falling quasistatic test particle.}
\nonumber
\end{equation} 
%

     {\it Gravitomagnetic field} 
$\vec{B}_{\rm g} :$
\begin{equation} 
\Omega_{\, \hat{i}}^{\rm gyro}  
\equiv - \frac{1}{2}  B_{\, \hat{i}}^{\rm g} \quad
\mbox{precession of gyro comoving with FIDO.}
\nonumber 
\end{equation}
%

    {\it Gravitomagnetic vector potential} 
$\vec{A}_{\rm g}:$ 
Because all 3-scalars must be unperturbed in the vector sector,
div $\vec{A}_{\rm g} \equiv 0,$ 
and $\vec{A}_{\rm g}$ is uniquely determined by $\vec{B}_{\rm g},$
\begin{equation} 
\vec{B}_{\rm g} =: \mbox{curl} \, \vec{A}_{\rm g} \, \, \, 
\Rightarrow  \, \, \, 
\vec{A}_{\rm g} \, = \, \vec{\beta} 
\, \equiv \,   
    \mbox{shift vector}.
\end{equation}
%

    {\it Our choice of FIDOs}: 
Our FIDOs are at fixed values of the spatial coordinates $x^i,$
and the spatial axes are fixed 
in the direction of our coordinate basis vectors.


\subsection{Einstein's $G^{\, \hat{0}}_{\, \, \hat{k}}$ Equation: 
The Momentum Constraint}


New result: The momentum constraint is 
    {\it form-identical} 
for 
    {\it all three} 
FRW background geometries, 
                 $K = (0, \pm 1):$
\begin{equation}
(- \Delta + \mu^2) \,   \vec{A}_{\, \rm g} 
= - 16 \pi G_{\rm N} \, \vec{J}_{\, \varepsilon}, 
\nonumber
\label{momentum.constraint}
\end{equation}
where 
    $\, (\mu/2)^2 \, \equiv \, - (dH/dt) \, 
    \equiv \, (H\mbox{-dot radius})^{-2},$
and 
    $\vec{J}_{\varepsilon} =$ {\it energy current density} =     
    {\it momentum density.}
Since the source in 
    Eq.~(\ref{momentum.constraint}) 
is the momentum density, this equation is called the
     {\it  'momentum constraint'.}

The momentum constraint is an    
    {\it elliptic equation,} 
i.e. there are 
    {\it no partial time-derivatives} 
of perturbations, 
although the momentum constraint refers to  
    {\it time-dependent} 
gravitomagnetism.

Our new approach: 
For the source we have used the 
    {\it LONB components}
    $\vec{J}^{\, \varepsilon}_{\, \hat{k}} 
     = T^{\hat{0}}_{\, \hat{k}},$ 
which is 
a
    {\it measurable input,} 
and 
which needs  
    {\it no prior knowledge} 
of $g_{0i},$ which is the output.
Einstein had emphasized that 
the coordinate-basis components $T^0_k$ are 
    {\it not} 
a directly measurable input:
    {\it 'If you have $T_{\mu \nu}$ and not a metric,
    the statement that matter by itself determines the metric 
    is meaningless.'} 
 
New result: 
The momentum constraint for time-dependent 
gravitomagnetism for all three FRW background geometries 
has the same form as Amp\`ere's law 
for stationary magnetism, 
except for the term $ \mu^2    \vec{A}_{\rm g},$
which causes causes a              
    {\it Yukawa suppression} 
beyond the 
    {\it $H$-dot radius.}
There are no curvature terms in 
    Eq.~(\ref{momentum.constraint}). 


\subsection{The Laplacian on Vector Fields in Riemannian 3-Spaces}


The Laplacian $\Delta$ acting on vector fields in 
    Eq.~(\ref{momentum.constraint}) 
is the  
   {\it de Rham - Hodge Laplacian,}
which mathematicians simply call 'the Laplacian',
and which differs from $\nabla^2,$ which mathematicians call 
the 'rough Laplacian'.
Unfortunately all publications on cosmological vector perturbations
up to ours have used the 'rough Laplacian'  $\nabla^2.$
The difference between the two operators is given by the
   {\it Weitzenb\"ock formula}: 
\begin{equation}
(\Delta - \nabla^2) \vec{A} 
              = - (2K/a_{\, \rm c}^{\, 2}) \vec{A},
\end{equation}
where $K = (\pm 1, 0)$ is the curvature index for the FRW background,
and $a_{\, \rm c}$ is its curvature radius.
For vorticity fields (divergence zero) the de Rham - Hodge Laplacian
is defined by   
\begin{equation}
(\Delta \, \vec{a})_{\mu} 
= - (\, {\rm curl \, curl} \, \vec{a}  \,)_{\mu}
= - (\, \star \, d \star d \, \tilde{a}\,)_{\mu},
\nonumber
\end{equation}
where we have given both the notation of elementary vector calculus
and the notation of calculus of differential forms with
$d \equiv $ exterior derivative and 
$\star \equiv $ Hodge dual. 

The de Rham - Hodge Laplacian on vector fields is 
      {\it singled out}
by the following properties:
\begin{enumerate}
\item     If all sources (curl and div) are zero 
          $\Rightarrow$ the de Rham - Hodge Laplacian gives zero.
\item     The de Rham - Hodge Laplcian 
                {\it commutes} 
          with curl, div, grad.
\item     The 
                {\it identities} 
          of vector calculus in Euclidean 3-space
          (familiar from Classical Electrodynamics) 
          remain true in Riemannian 3-spaces
          for the Hodge - de Rham Laplacian.
\item     The 
                {\it action principle} 
          for Amp\`ere magnetism in Riemannian 3-spaces
          directly produces Amp\`ere's equation for $\vec{A}$
          with the Hodge - de Rham Laplacian and 
                {\it without curvature terms}.
\item     For electromagnetism in curved space-time the 
                {\it equivalence principle forbids curvature terms}
          in equations with the Hodge - de Rham Laplacian.
\end{enumerate}
Every one of these properties does not hold 
for the 'rough' Laplacian $\nabla^2.$


\section{The Bottom Lines}

\subsection{The Solution of the Momentum Constraint}


Cosmological gravitomagnetism on a background of
               {\it open} 
FRW universes gives
               {\it identical expressions} 
for $ K = 0$ and for $K = -1:$
%
\begin{eqnarray}
 \vec{B}_{{\rm g}} (P) &=& -2 \, \vec{\Omega}_{\rm gyro} (P) =
\nonumber
\label{solution.momentum.constraint}
\\
&=& -4 \, G_{\rm N} \int d({\rm vol}_Q) \,  
[\vec{n}_{PQ} \times \vec{J}_{\varepsilon}(Q)] \, 
 Y_{\mu} (r_{PQ})
\\
 Y_{\mu} (r) 
&=& 
\frac{-d}{dr} \, [\frac{1}{R} \exp (- \mu r)] 
= {\rm Yukawa \, \, force},
\label{Yukawa.force}
\end{eqnarray}
where $r =$ radial distance,
and $2 \pi R =$ circumference of the great circle 
through $Q$ and centered at $P.$
Vectors are parallel-transported from $Q$ to $P$ 
along the connecting geodesic.~---
The solution 
   Eq.~(\ref{solution.momentum.constraint}) 
is analogous to Amp\`ere's solution for stationary magnetism,
but 
   Eq.~(\ref{solution.momentum.constraint})
is valid for 
   {\it time-dependent} 
gravito-magnetodynamics, and it has a 
   {\it Yukawa suppression.} 


There is a fundamental difference
between our solution 
    Eq.~(\ref{solution.momentum.constraint})
for cosmological gravitomagnetism
and the corresponding solutions in other theories,
Amp\`ere's magnetism, electromagnetism in Minkowski space, 
and General Relativity in the solar system: 
Our solution for Cosmological General Relativity is manifestly 
   {\it form-invariant} 
when going to globally rotating frames,
while the 
   {\it solutions} 
in other theories are 
   {\it not}
form-invariant
when going to globally rotating frames.


If the background is a 
   {\it closed} FRW universe, 
one makes the following replacement in   
    Eq.~(\ref{Yukawa.force}):
%
\begin{equation}
\exp(- \mu r) \, \, \, \, \Rightarrow \, \, \, \, 
\sinh^{-1}(\mu \pi) \,  \sinh[\mu(\pi - r)].
\label{replacement}
\end{equation}
%


\subsection{Exact Dragging of Inertial Axes}


From 
    {\it symmetry} 
under rotations and reflections
one concludes: 
The precession of a gyroscope
can only be acted on 
by the component of the matter velocity field 
in the vorticity sector (not by scalar or tensor perturbations)
and with $J^P = 1^+$ relative to the gyroscope.
This component of the velocity field 
is equivalent to a rigid rotation of matter  
with angular velocity $\vec{\Omega}_{\rm matter} (r).$

From
    Eq.~(\ref{solution.momentum.constraint})
one concludes that inertial axes, i.e. the spin axes of 
     {\it gyroscopes}, 
     {\it exactly follow}
the 
     {\it weighted average} 
of the energy currents of cosmic 
     {\it matter},  
\begin{eqnarray}
\vec{\Omega}_{\rm gyro} 
&=& <\vec{\Omega}_{\rm matter}> \,  \, \, \,  
\equiv \int_0^\infty  dr 
\, \, \vec{\Omega}_{\rm matter}(r) \, \, W(r)   
\\
W(r) 
&=& \frac{1}{3} \, 16 \pi G_{\rm N} (\rho +p) \, R^3 \, Y_{\mu}(r),
\end{eqnarray}
for perturbations of 
open FRW universes.   
The 
    {\it weight function} 
$W(r)$  is 
    {\it normalized} 
to   
    {\it unity,}
\begin{equation}
\int_0^\infty dr \, W(r) = 1,
\end{equation}
as it must be for a proper averageing weight function 
in any problem.~---
For perturbations of 
a closed FRW universe   
one again makes the replacement of
    Eq.~(\ref{replacement}).



\end{document}